\documentclass[pdflatex,sn-mathphys-num]{sn-jnl}

\usepackage{graphicx}%
\usepackage{multirow}%
\usepackage{amsmath,amssymb,amsfonts}%
\usepackage{amsthm}%
\usepackage{mathrsfs}%
\usepackage[title]{appendix}%
\usepackage{xcolor}%
\usepackage{textcomp}%
\usepackage{manyfoot}%
\usepackage{booktabs}%
\usepackage{algorithm}%
\usepackage{algorithmicx}%
\usepackage{algpseudocode}%
\usepackage{listings}%
\usepackage{array}
\usepackage{pifont}
\usepackage{booktabs}
\newcolumntype{P}[1]{>{\raggedright\arraybackslash}p{#1}}

\theoremstyle{thmstyleone}%
\theoremstyle{thmstyletwo}%

\theoremstyle{thmstylethree}%

\begin{document}

\title[Article Title]{NeuroNarrator: A Generalist EEG-to-Text Foundation Model for Clinical Interpretation via Spectro-Spatial Grounding and Temporal State-Space Reasoning}


\author[1]{\fnm{Guoan} \sur{Wang}}\email{gwang31@stevens.edu}
\equalcont{These authors contributed equally to this work.}

\author[1]{\fnm{Shihao} \sur{Yang}}\email{syang57@stevens.edu}
\equalcont{These authors contributed equally to this work.}

\author[1]{\fnm{Jun-en} \sur{Ding}}\email{jding17@stevens.edu}

\author[1]{\fnm{Hao} \sur{Zhu}}\email{hzhu38@stevens.edu}

\author*[1]{\fnm{Feng} \sur{Liu}}\email{fliu22@stevens.edu}

\affil*[1]{\orgdiv{Department of Systems Engineering}, \orgname{Stevens Institute of Technology}, \orgaddress{\street{1 Castle Point Terrace}, \city{Hoboken}, \postcode{07030}, \state{New Jersey}, \country{USA}}}


\abstract{Electroencephalography (EEG) provides a non-invasive window into neural dynamics at high temporal resolution and plays a pivotal role in clinical neuroscience research. Despite this potential, prevailing computational approaches to EEG analysis remain largely confined to task-specific classification objectives or coarse-grained pattern recognition, offering limited support for clinically meaningful interpretation. To address these limitations, we introduce NeuroNarrator, the first generalist EEG-to-text foundation model designed to translate electrophysiological segments into precise clinical narratives. A cornerstone of this framework is the curation of NeuroCorpus-160K, the first harmonized large-scale resource pairing over 160,000 EEG segments with structured, clinically grounded natural-language descriptions. Our architecture first aligns temporal EEG waveforms with spatial topographic maps via a rigorous contrastive objective, establishing spectro–spatially grounded representations. Building on this grounding, we condition a Large Language Model through a state-space–inspired formulation that integrates historical temporal and spectral context to support coherent clinical narrative generation. This approach establishes a principled bridge between continuous signal dynamics and discrete clinical language, enabling interpretable narrative generation that facilitates expert interpretation and supports clinical reporting workflows. Extensive evaluations across diverse benchmarks and zero-shot transfer tasks highlight NeuroNarrator’s capacity to integrate temporal, spectral, and spatial dynamics, positioning it as a foundational framework for time–frequency–aware, open-ended clinical interpretation of electrophysiological data.}




\maketitle
\section{Introduction}

Electroencephalography (EEG) provides a non-invasive window into human brain activity, enabling high–temporal-resolution measurement of neural dynamics at the millisecond scale\cite{michel2019eeg, he2018electrophysiological}. Compared to other neuroimaging modalities such as functional magnetic resonance imaging (fMRI)\cite{logothetis2008we}, magnetoencephalography (MEG)\cite{baillet2017magnetoencephalography}, or intracranial EEG (iEEG)\cite{parvizi2018promises}, EEG offers a favorable trade-off among accessibility, temporal resolution, and cost, making it widely adopted across clinical diagnostics, cognitive neuroscience, and brain–computer interface (BCI) applications\cite{mcfarland2011brain}. As a result, EEG has been extensively investigated in a wide range of downstream tasks, including seizure epilepsy classification\cite{boonyakitanont2020review}, motor imagery recognition\cite{amin2019deep}, sleep stage classification\cite{supratak2017deepsleepnet}, and cognitive workload assessment\cite{chikhi2022eeg}. 

With the rapid advancement of deep learning, neural network–based approaches have been widely adopted for EEG analysis, yielding substantial performance gains across a variety of downstream tasks. Despite these advances, most EEG deep learning models are still optimized for narrowly defined downstream objectives\cite{schirrmeister2017deep, lawhern2018eegnet}. Although recent foundation model efforts have begun to leverage large-scale pretraining\cite{yang2023biot}, EEG-to-text interpretation remains underexplored beyond rigid formulations constrained by labels or templates, particularly in settings that require open vocabulary and free-form clinical descriptions\cite{jiang2024neurolm}. 
In parallel, recent studies have invesitgated EEG-to-text interpretation in decoding-oriented settings, where textual targets correspond to external stimuli, experimental instructions, or subjects' linguistic outputs\cite{duan2023dewave}. However, clinical EEG interpretation does not aim to decode latent semantic content from neural activity, but to describe and contextualize electrophysiological patterns as interpreted by clinical experts. Moreover, existing studies that link EEG recordings to clinical narratives typically operate at the recording-level, overlooking the fine-grained temporal evolution that underlies clinically meaningful electrographic dynamics and thereby limiting their ability to capture and interpret relevant biomarkers\cite{ndir2025eeg, yin2025neurolexlightweightdomainlanguage}.

Motivated by the requirements of clinical EEG reporting, where diagnostically relevant EEG features are often transient, sparse, and should be described with explicit time anchors\cite{kane2017revised, 775d5bc3540a43fdaefa6f4dfcca7f24}, we pose EEG-to-text interpretation at the level of short segments rather than entire recording. Temporal localization and time-resolved interpretation are essential, because recording-level generation can blur transient electrographic events and disrupt the temporal evidence needed for clinician verification\cite{herman2015consensus}. Beyond event detection, we formulate EEG-to-text interpretation as the generation of open-vocabulary, segment-level descriptions that explicitly preserve waveform morphology, spectral structure, and spatial topography—core elements of expert clinical reading—and thereby support compositional generalization across diverse electrophysiological patterns. Such time-resolved, signal-grounded narratives are intended to support workflow by directing attention to suspicious epochs for rapid review and by facilitating standardized documentation, thereby reducing reporting burden while leaving final adjudication to experts.


\begin{figure}[t]
  \centering
  \includegraphics[width=\textwidth]{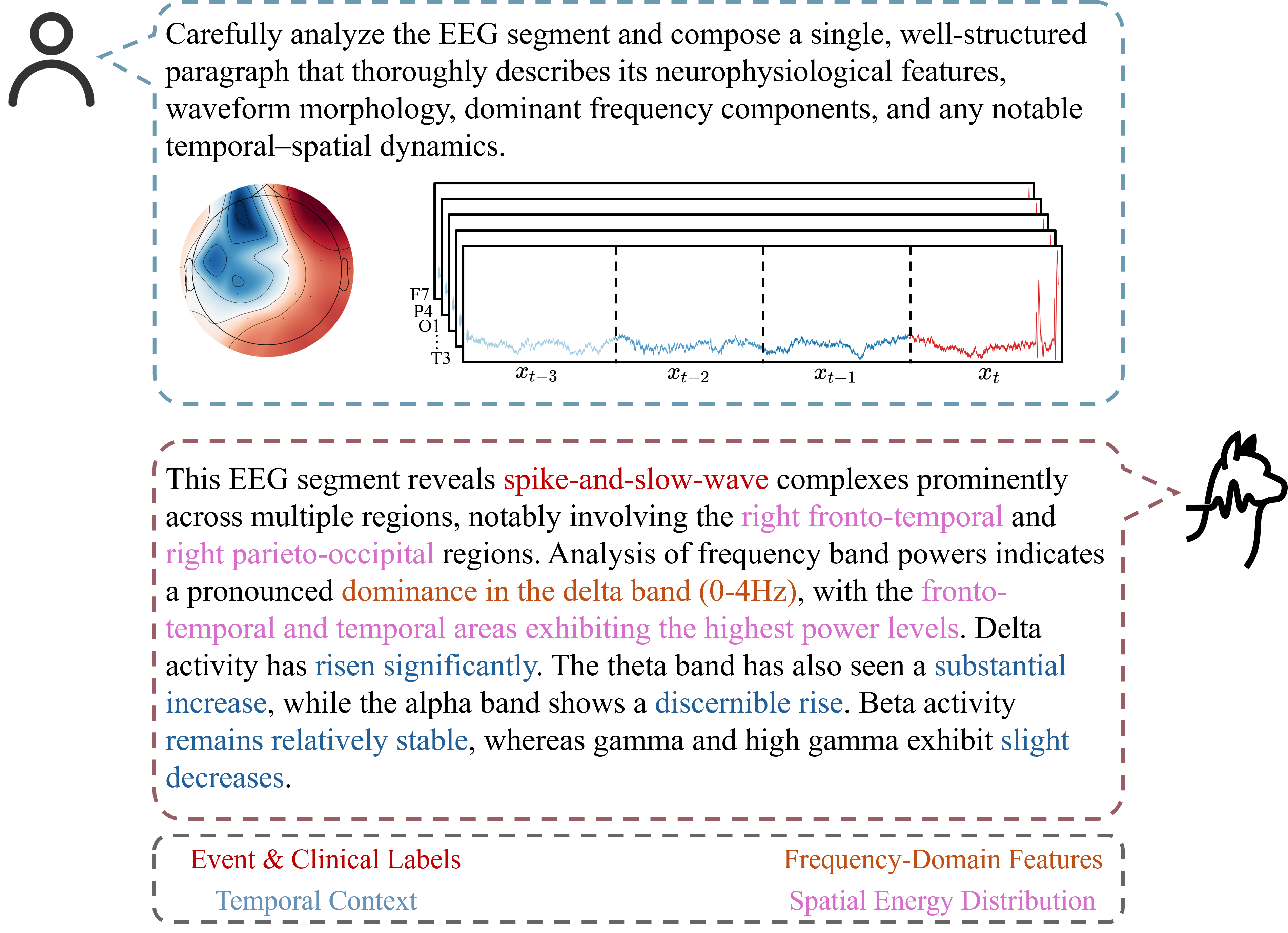}%
  \caption{\textbf{Illustrative example of segment-level, clinically grounded EEG interpretation.}
In contrast to coarse recording-level interpretaion, this sample demonstrates the generation of a fine-grained clinical narrative for a 10-second segment. The generated text systematically synthesizes four dimensions of electrophysiological analysis: (i) Event \& Clinical Labels, identifying specific morphological abnormalities (e.g., spike-and-slow-wave complexes); (ii) Spatial Energy Distribution, capturing prominent signal and energy variations in specific anatomical regions (e.g., right fronto-temporal); (iii) Frequency-Domain Features, identifying the dominant spectral bands; and (iv) Temporal Context, characterizing the non-stationary evolution of brain states relative to preceding segments.}
  \label{fig:example}
\end{figure}

In this work, we introduce \textbf{NeuroNarrator}, the first generalist EEG-to-text foundation model designed to bridge electrophysiological recordings and clinically meaningful natural-language interpretation, as exemplified in Fig \ref{fig:example}. By moving beyond coarse recording-level descriptions, our framework explicitly preserves the fine-grained spectro-spatial structure of transient EEG patterns and models their evolution over time, enabling linguistically grounded narratives that reflect both localized electrophysiological features and temporally evolving brain states.

To achieve this, we implement a unified framework that synergizes large-scale open-vocabulary representation learning with rigorous spectro-spatial grounding. By integrating these representations within a state-space formulation, we bridge the semantic gap between electrophysiology and clinical narrative through the following core contributions:

\begin{enumerate}
  \item \textbf{Construction of NeuroCorpus-160K.} We aggregate and standardize 16 heterogeneous datasets into the first large-scale, open-vocabulary EEG-clinical narrative corpus. To ensure robust generalization, we establish a rigorous subject-disjoint training and evaluation split, providing a standardized benchmark for open-vocabulary interpretation beyond closed-set classification.
  \item \textbf{Contrastive Spectro-Spatial Alignment.} We introduce a multimodal alignment mechanism that projects temporal EEG waveforms and spatial topographic maps into a shared semantic manifold. This enforces strict correspondence between spectral dynamics and spatial energy distributions, resolving the grounding ambiguity inherent in single-modality encoding.
  \item \textbf{Unified State-Space Generalist Framework.} We propose NeuroNarrator, a multimodal large language model (MLLM) architecture that integrates spectro-spatial features with state-space-inspired temporal-spectral priors. By conditioning generation on historical trajectory embeddings, the framework captures evolving brain dynamics (e.g., seizure evolution) within an end-to-end decoding process.
\end{enumerate}

\section{Related Work}
\subsection{Generalist EEG Representation Learning}
The landscape of automated EEG analysis has undergone a fundamental transition, shifting from task-specific supervised architectures trained on limited datasets to generalist foundation models pre-trained on massive, heterogeneous corpora via self-supervision. Pioneering efforts, such as BENDR\cite{kostas2021bendr}, adapted contrastive learning objectives from speech processing (e.g., wav2vec 2.0) to extract transferable representations directly from EEG sequences. Building on this, recent frameworks have increasingly leveraged Masked Signal Modeling (MSM) to scale pre-training efficacy. Notably, LaBraM\cite{jiang2024large} introduces vector-quantized neural spectrum prediction, enabling the model to learn generic spectral features by reconstructing masked frequency patches. Concurrently, BIOT\cite{yang2023biot} addresses the challenge of cross-dataset heterogeneity by tokenizing channels independently, thereby allowing the model to generalize across varying electrode montages and sampling rates.

To further resolve the complex spatiotemporal dynamics of brain activity, specialized attention mechanisms have been proposed. EEGPT\cite{wang2024eegpt} employs large-scale transformers to extract hierarchical temporal features from extensive scalp EEG data, while CBraMod\cite{wang2024cbramod} utilizes a ``criss-cross'' attention mechanism to explicitly decouple spatial and temporal dependencies, enhancing structural encoding stability. However, a critical limitation persists: these frameworks encode EEG signals into abstract latent representations that lack explicit semantic grounding. By treating brain activity strictly as numerical time-series rather than clinically meaningful physiological events, they remain disconnected from the open-ended linguistic reasoning required for expert-level interpretation.

\subsection{EEG-to-Text Generation and Multimodal Alignment}
While vision-language models have established a robust paradigm for grounding visual semantics in natural language, bridging the representational gap between electrophysiology and clinical linguistics remains a nascent frontier. Existing efforts in this domain have predominantly diverged into two streams. The first stream focuses on ``brain-to-text'' decoding, which aims to reconstruct external stimuli, experimental prompts, or the subject’s internal linguistic content from neural activity. For instance, DeWave\cite{duan2023dewave}introduces discrete codex encoding to translate neural dynamics into open-vocabulary text, while strictly discriminative approaches have explored similar alignment strategies for sentiment classification\cite{wang2022open} or cross-lingual decoding\cite{lu2025eeg2text}. However, clinical interpretation entails a fundamentally different objective: it seeks not to decode latent cognitive content, but to articulate precise descriptions of the electrophysiological phenomena themselves.

The second stream, which attempts to link EEG recordings to clinical narratives, currently faces significant limitations in granularity and generative flexibility. Alignment-centric frameworks, such as EEG-CLIP\cite{ndir2025eeg} and recent EEG-language pretraining frameworks\cite{pmlr-v267-gijsen25a}, primarily leverage cross-modal contrastive learning for retrieval tasks or coarse clinical phenotyping, lacking the capacity for fine-grained descriptive generation. Conversely, generative models like NeuroLex\cite{yin2025neurolexlightweightdomainlanguage} typically operate at the macroscopic recording level. This coarse resolution inherently obscures the fine-grained temporal structure of EEG, failing to ground transient pathological morphologies in precise linguistic descriptions. Similarly, generalist architectures like NeuroLM\cite{jiang2024neurolm} rely heavily on rigid instruction templates to map signals to task outputs. While effective for closed-set execution, this paradigm restricts the compositional generalization requisite for open-ended interpretation, where a model must synthesize evolving spectro-spatial dynamics into a coherent, non-templated narrative.

\subsection{Spectro-Spatial and Temporal Modeling in Neuroscience}
Expert clinical EEG interpretation relies on the integration of spectral and topographic evidence, rather than the analysis of isolated time-series. The semantic validity of an oscillatory feature is determined by its spatial localization; for instance, alpha-band activity ($8\text{--}12$ Hz) constitutes a normative resting rhythm when posteriorly dominant, yet indicates pathology (e.g., ``alpha coma'') when generalized or anteriorly distributed\cite{srinivasan2007eeg, buzsaki2004neuronal}. This interdependence aligns with EEG microstate theory, which posits that global brain states manifest as quasi-stable topographic potential maps distinct from purely temporal frequency features\cite{michel2018eeg}. Consequently, accurate machine interpretation requires a rigorous coupling of spectral content with spatial topology.

Furthermore, neural activity is governed by non-stationary dynamical systems, where instantaneous observations are conditioned on the trajectory of evolving latent states\cite{breakspear2017dynamic}. Pathological phenomena, such as seizure propagation or shifts in consciousness, are not discrete static events but continuous transitions through a latent state space\cite{jirsa2017virtual}. However, prevailing deep learning architectures often neglect these neurodynamical principles by flattening spatial topology into generic channel vectors and processing segments as independent snapshots. By failing to model the conditional dependence of the current segment on its historical trajectory, these ``black-box'' approaches lack the state-space reasoning required to generate narratives that reflect the temporal continuity characteristic of expert clinical reports.

\section{Methods}
\subsection{Overview of the NeuroNarrator Framework}

\begin{figure}[t]
  \centering
  \includegraphics[width=\textwidth]{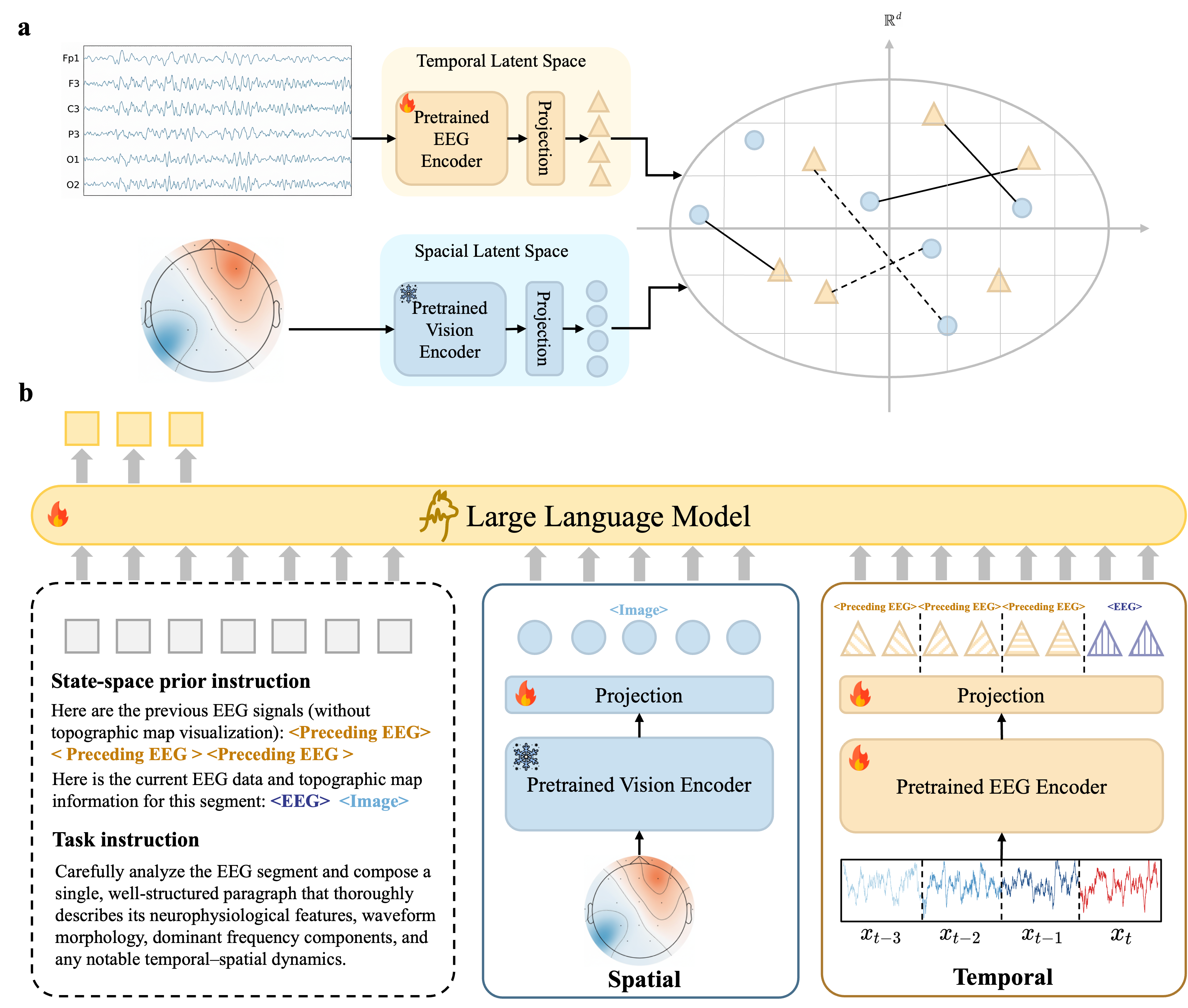}%
  \caption{\textbf{NeuroNarrator architecture for spectro–spatially grounded and temporally coherent EEG-to-text generation.}
(a) Dual-stream spectro–spatial grounding encodes each EEG segment using a pretrained EEG encoder operating on multichannel waveforms and a frozen vision encoder processing the corresponding scalp topographic map. Modality-specific features are projected into a shared latent space and aligned via a contrastive objective, enforcing correspondence between spectral dynamics and spatial energy distributions. (b) State-space–inspired generative modeling conditions text generation on both the aligned spectro–spatial embedding of the current segment and a short trajectory of preceding EEG segments, serving as a proxy for latent brain-state evolution. These continuous embeddings are injected as soft prompt tokens, replacing designated placeholder positions in the language-model prompt alongside task instructions, enabling the synthesis of clinically grounded narratives that preserve waveform morphology, dominant frequency structure, spatial localization, and temporal dynamics.
}
\label{fig:model}
\end{figure}

NeuroNarrator is a unified MLLM for segment-level EEG-to-text interpretation, designed to bridge electrophysiological recordings and clinically meaningful natural-language descriptions. The framework integrates complementary temporal and spatial representations of EEG signals and incorporates short-term temporal context to support coherent clinical narration. An overview of the model architecture and data flow is provided in Fig. \ref{fig:model}, with detailed components and learning objectives described in the following sections.

\subsection{Construction of a Clinically Grounded EEG–Text Corpus}
\subsubsection{Multi-Source EEG Data Harmonization}
\begin{figure}[t]
  \centering
  \includegraphics[width=\textwidth]{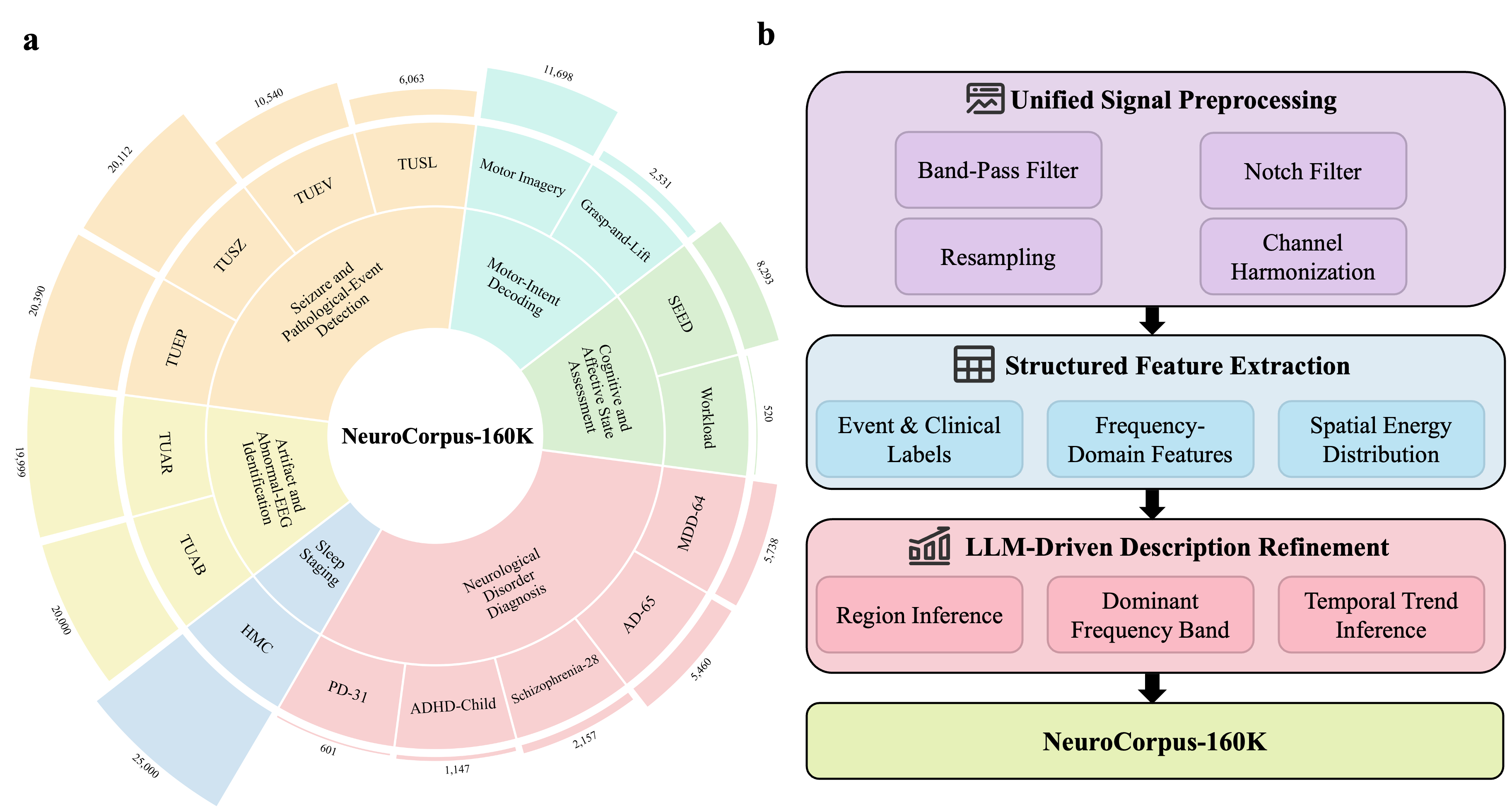}%
  \caption{\textbf{Overview of NeuroCorpus-160K construction.}
(a) Distribution of the aggregated datasets across major clinical domains.
(b) The unified data processing workflow, which transforms raw recordings into clinically grounded narratives via three stages: signal preprocessing, structured feature extraction, and LLM-driven description refinement.}
  \label{fig:dataset}
\end{figure}

To systematically address the electrophysiological heterogeneity encountered in clinical environments, we constructed NeuroCorpus-160K by harmonizing 16 publicly available datasets across diverse acquisition hardware and protocols (Table \ref{tab:eeg_datasets}). As visualized in Fig. \ref{fig:dataset}a, this collection covers the primary clinical application domains of neurological disorder diagnosis, seizure and pathological-event detection, artifact and abnormal-EEG identification, cognitive and affective state assessment, motor-intent decoding, and sleep staging. Such phenotypic and instrumental diversity is critical for mitigating single-source biases and ensuring the model captures a broad spectrum of signal morphologies.

\begin{table}[htbp]
\centering
\caption{\textbf{Constituent datasets and split statistics for NeuroCorpus-160K.}
Summary of the 16 public EEG datasets integrated into NeuroCorpus-160K, including the associated task and the sizes of the training and held-out test splits, reported as the number of non-overlapping 10-s segments (Samples) and total recording duration (Hours).}
\label{tab:eeg_datasets}
\begin{tabular}{@{}llrrrr@{}}
\toprule
\multirow{2}{*}{\textbf{Dataset}} & \multirow{2}{*}{\textbf{Task}} & \multicolumn{2}{c}{\textbf{Train}} & \multicolumn{2}{c}{\textbf{Test}} \\
\cmidrule(lr){3-4} \cmidrule(lr){5-6}
 &  & \textbf{Samples} & \textbf{Hours} & \textbf{Samples} & \textbf{Hours} \\
\midrule
AD-65\cite{ds004504:1.0.2} & Alzheimer's Disease Detection & 5,460 & 15.17 & 1,382 & 3.84 \\
ADHD-Child\cite{rzfh-zn36-20} & ADHD Detection & 1,147 & 3.19 & 330 & 0.92 \\
Grasp-and-Lift\cite{luciw2014multi} & Hand Manipulation Event Detection & 2,531 & 7.03 & 894 & 2.48 \\
HMC\cite{alvarez2021inter} & Sleep Stage Classification & 25,000 & 69.44 & 1,000 & 2.78 \\
MDD-64\cite{Mumtaz2016} & Major Depressive Disorder Detection & 5,738 & 15.94 & 1,466 & 4.07 \\
Motor Imagery\cite{schalk2004bci2000} & Motor Imagery Recognition & 11,698 & 32.49 & 1,191 & 3.31 \\
PD31\cite{ds002778:1.0.5} & Parkinson's Disease Analysis & 601 & 1.67 & 217 & 0.60 \\
Schizophrenia-28\cite{repod.0107441_2017} & Schizophrenia Detection & 2,157 & 5.99 & 680 & 1.89 \\
SEED\cite{zheng2015investigating} & Emotion Recognition & 8,293 & 23.04 & 1,050 & 2.92 \\
TUAB\cite{harati2015improved} & Abnormal EEG Detection & 20,000 & 55.56 & 1,000 & 2.78 \\
TUAR\cite{buckwalter2021recent} & Artifact Classification & 19,999 & 55.55 & 865 & 2.40 \\
TUEP\cite{veloso2017big} & Epilepsy Detection & 20,390 & 56.64 & 1,000 & 2.78 \\
TUEV\cite{harati2015improved} & Event-Type Classification & 10,540 & 29.28 & 1,000 & 2.78 \\
TUSL\cite{von2017electroencephalographic} & Slowing-Event Classification & 6,063 & 16.84 & 549 & 1.52 \\
TUSZ\cite{shah2018temple} & Seizure Event Detection & 20,112 & 55.87 & 1,000 & 2.78 \\
Workload\cite{zyma2019electroencephalograms} & Mental Workload Assessment & 520 & 1.44 & 90 & 0.25 \\
\midrule
\textbf{Total} & \textbf{16 Tasks} & \textbf{160,249} & \textbf{445.14} & \textbf{13,714} & \textbf{38.09} \\
\bottomrule
\end{tabular}
\end{table}

The consolidated NeuroCorpus-160K comprises a total of 483.23 hours of recordings from subjects ranging from early childhood to advanced age. To facilitate rigorous benchmarking, we partitioned the entire corpus at the subject level into a training set (160,249 segments) and a strictly separated held-out evaluation subset (13,714 segments). Detailed split criteria are provided in Section \ref{sec_datasets}.

\subsubsection{Unified Signal Preprocessing}\label{lab:sec3.2}

To mitigate the distributional shifts introduced by varying acquisition hardware and clinical protocols, we implement a rigorous signal standardization pipeline across all constituent datasets of NeuroCorpus-160K, as illustrated in Fig. \ref{fig:dataset}b. We first apply a zero-phase finite impulse response (FIR) band-pass filter ($0.1\text{--}75\text{ Hz}$) to all recordings to isolate clinically relevant spectral content while attenuating low-frequency drift and high-frequency noise. This is followed by a region-specific notch filter ($50\text{ Hz}$ or $60\text{ Hz}$) to suppress power-line interference. To ensure temporal uniformity across the heterogeneous corpus, all time-series are strictly resampled to a fixed sampling rate of $200\text{ Hz}$. Electrode montages were strictly harmonized to the international 10–20 or 10–10 systems to ensure spatial consistency.

\subsubsection{Structured Feature Extraction and LLM-Driven Description Refinement} \label{lab:sec3.3}
To systematically bridge the representational gap between continuous electrophysiological dynamics and discrete linguistic descriptions, we implemented a unified pipeline that transforms preprocessed EEG signals into high-fidelity clinical narratives(Fig. \ref{fig:dataset}b). Following signal preprocessing, the continuous recordings were partitioned into non-overlapping 10-second segments ($x \in \mathbb{R}^{C \times T}$). For each segment, we constructed a structured quantitative feature template to serve as a factual scaffold for the subsequent generation process. This template explicitly encodes three complementary dimensions: (i) Event \& Clinical Labels, capturing subject-level clinical context (e.g., demographics, diagnosis, task/condition, and dataset-provided event annotations when available); (ii) Frequency-Domain Features, derived from power spectral density (PSD) estimates summarized over canonical frequency bands; and (iii) Spatial Energy Distribution, derived by computing channel-wise power and partitioning electrodes into three distinct intensity tiers via K-means clustering ($K=3$). 

\begin{table*}[t]
  \centering
  \small
  \caption{\textbf{Example of the transformation from structured quantitative scaffolds to clinically grounded narratives.}
  This comparison demonstrates the synthesis of rigid feature templates (left) into a fluent, coherent description (right). Note that while the Frequency-Domain refinement is conditioned on the spectral history of preceding segments ($\{x_{t-k}\}_{k=1}^N$) alongside the current segment ($x_t$) to infer temporal trends, only the current segment’s features are listed here for conciseness.}
  \label{tab:caption_pipeline}
  \begin{tabular}{@{}p{0.28\textwidth}p{0.35\textwidth}p{0.38\textwidth}@{}}
    \toprule
    \textbf{Component} & \textbf{Structured Template} & \textbf{LLM Refined Descriptions} \\
    \midrule

    \textbf{Event \& Clinical Labels}
    & This EEG segment from a 19-year-old female subject is recorded during a mental arithmetic task with cognitive load.
    & In this EEG segment from a 19-year-old female subject recorded during a mental arithmetic task with cognitive load, the observed patterns reflect task-related dynamics across frequency bands and scalp regions. \\
    \addlinespace[0.6em]

    \textbf{Frequency-Domain Features(Conditioned on $\{x_{t-k}\}_{k=1}^{N}, x_t$)}
    & Frequency band powers: delta (0--4~Hz): 39.8676~$\mu\mathrm{V}^2$; theta (4--8~Hz): 11.5915~$\mu\mathrm{V}^2$; alpha (8--12~Hz): 5.7254~$\mu\mathrm{V}^2$; beta (12--30~Hz): 5.2747~$\mu\mathrm{V}^2$; gamma (30--50~Hz): 1.1490~$\mu\mathrm{V}^2$; high-gamma (50--75~Hz): 0.0810~$\mu\mathrm{V}^2$.
    & The delta frequency band (0--4~Hz) exhibits the highest power and shows a moderate decrease. Theta activity (4--8~Hz) shows a noticeable increase, while the alpha band (8--12~Hz) presents a significant decrease. Stable readings are observed in both the beta (12--30~Hz) and gamma (30--50~Hz) bands, while the high-gamma band (50--75~Hz) remains consistently low. \\
    \addlinespace[0.6em]

    \textbf{Spatial Energy Distribution}
    & Highest-energy channel is C4 (103.2591~$\mu\mathrm{V}^2$). High-energy channels: FZ, C4, PZ. Medium-energy channels: F3, F4, F8, C3, CZ, P3, P4, T4, T6. Low-energy channels: F7, O1, O2, T3, T5.
    & The EEG power is predominantly concentrated in the fronto-central and parietal regions, with the C4 channel maintaining the highest energy levels. \\
    \addlinespace[0.6em]

    \bottomrule
  \end{tabular}
\end{table*}

These structured yet rigid feature representations were subsequently refined into fluent, clinically coherent narratives using GPT-4.1\cite{achiam2023gpt}. By employing a domain-specific prompting strategy, the model was directed to synthesize the quantitative template into a natural language description, performing complex inference tasks such as identifying dominant anatomical regions (e.g., right fronto-temporal) and characterizing the primary oscillatory modes. Additionally, we incorporated implicit temporal context from preceding segments to allow the model to infer and describe evolving dynamic trends, such as the gradual buildup of theta rhythm or the attenuation of alpha activity, reflecting the non-stationary nature of brain states. 
Table \ref{tab:caption_pipeline} exemplifies the transformation from structured feature templates to final refined descriptions.

\subsubsection{State-Space-Inspired Temporal Context Modeling}
Clinically significant electrophysiological phenomena, including the gradual evolution of a seizure, the onset of drowsiness, or shifts in cognitive workload, are inherently non-stationary. These dynamics are best characterized not as a series of independent snapshots, but as continuous trajectories through a latent manifold of brain states. To capture this temporal continuity within NeuroCorpus-160K, we formulate the clinically grounded corpus generation process using a state-space perspective.

Formally, let the sequence of EEG segments be denoted by $\{x_{t}\}$, where each observation $x_t$ derives from a latent neural state $z_t$. In standard interpretation paradigms, the clinical description $y_t$ is typically generated via a direct mapping $x_t \to y_t$, effectively discarding historical context. However, the underlying physiological state evolves according to transition dynamics $z_t \approx \mathcal{F}(z_{t-1}, x_{t-1})$, implying that the accurate interpretation of the current segment $x_t$ is conditionally dependent on its trajectory.
To operationalize this, we approximate the latent trajectory by explicitly conditioning the target description for $x_t$ on a local temporal neighborhood. Specifically, we construct the data samples to include the preceding $N$ segments $\{x_{t-k}\}_{k=1}^{N}$ as a proxy for the system's history.

\subsection{Spectro-Spatial Representation Learning} \label{sec4.2}

\subsubsection{Dual-Stream Spectro-Spatial Encoding}
To capture the distinct yet complementary information inherent in electrophysiological data, we employ a dual-stream encoding architecture that processes temporal dynamics and spatial topology in parallel before projecting them into a shared latent space, as illustrated in Fig. \ref{fig:model}a.

\textbf{Temporal EEG Representation.} Let $x_t \in \mathbb{R}^{C \times T}$ denote the $t$-th preprocessed EEG segment, where $C$ represents the channel dimension and $T$ corresponds to the temporal window (e.g., 2000 time points at 200 Hz). To extract rich temporal dependencies, we utilize a domain-specific EEG encoder, $f_{\text{eeg}}(\theta^{\text{eeg}})$, instantiated as the LaBraM-Base\cite{jiang2024large} architecture. We initialize $f_{\text{eeg}}$ with weights pre-trained on large-scale EEG corpora via self-supervised learning, ensuring the model possesses a robust prior for canonical oscillatory patterns and artifact rejection. The encoder processes $x_t$ and aggregates the temporal token sequence via mean pooling to yield a compact latent representation $\mathbf{h}_t^{\text{eeg}}$:
\begin{equation}
\mathbf{h}_t^{\text{eeg}} = f_{\text{eeg}}(x_t \mid \theta^{\text{eeg}} ) \in \mathbb{R}^{d_1},
\end{equation}
where $d_1$ denotes the hidden dimensionality of the encoder.

\textbf{Spatial Topographic Representation.} Complementing the temporal stream, we explicitly model the spatial distribution of scalp potentials. Drawing inspiration from ConvDip\cite{hecker2021convdip}, which validated the efficacy of learning spatially organized patterns from topographic projections, we generate a corresponding EEG topographic map $i_t \in \mathbb{R}^{H \times W \times 3}$ for each segment $x_t$. We employ the CLIP ViT-Large\cite{radford2021learning} vision encoder, $f_{\text{vis}}(\theta_{\text{vis}})$, to process this map. Crucially, we keep $f_{\text{vis}}$ frozen to preserve its pre-trained vision-language alignment, thereby anchoring spatial features in a semantically rich manifold and compelling the EEG encoder to align with this stable, language-grounded target. We obtain a global spatial representation $\mathbf{h}_t^{\text{vis}}$ by averaging the output patch embeddings:
\begin{equation}
    \mathbf{h}_t^{\text{vis}} = f_{\text{vis}}(i_t \mid \theta_{\text{vis}}) \in \mathbb{R}^{d_2},
\end{equation}
where $d_2$ is the output dimensionality of the vision encoder.

\textbf{Projection to Shared Manifold.} To align these heterogeneous representations, we employ lightweight projection heads $g_{\text{EEG}}(\theta_{\text{proj}}^{\text{eeg}})$ and $g_{\text{vis}}(\theta_{\text{proj}}^{\text{vis}})$, implemented as two-layer MLPs with GELU activation. These projectors map the modality-specific features into a common $d$-dimensional embedding space:
\begin{equation}
    \mathbf{z}_t^{\text{eeg}} = g_{\text{eeg}}(\mathbf{h}_t^{\text{eeg}} \mid \theta_{\text{proj}}^{\text{eeg}}), \quad \mathbf{z}_t^{\text{vis}} = g_{\text{vis}}(\mathbf{h}_t^{\text{vis}} \mid \theta_{\text{proj}}^{\text{vis}}),
\end{equation}
where $\mathbf{z}_t^{\text{eeg}}, \mathbf{z}_t^{\text{vis}} \in \mathbb{R}^{d}$.

\subsubsection{Contrastive Spectro-Spatial Alignment}
To enforce a rigorous semantic correspondence between the temporal evolution of the EEG segment, $x_t$, and its instantaneous spatial organization, we employ a contrastive learning objective. Specifically, inspired by the optimization landscape of SigLIP\cite{zhai2023sigmoid}, we leverage a sigmoid-based strategy to decouple the normalization of positive and negative pairs.

Let $\hat{z}_i^{\text{eeg}}$ and $\hat{z}_j^{\text{vis}}$ denote the $L_2$-normalized embeddings of the $i$-th EEG segment and the $j$-th EEG topographic map within a mini-batch of size $B$. We define the pairwise similarity logit $s_{ij}$ between the temporal and spatial representations as:
\begin{equation}
    s_{ij} = \tau \cdot (\hat{z}_i^{\text{eeg}})^\top \hat{z}_j^{\text{vis}} + b,
\end{equation}
where $\tau$ is a learnable temperature parameter and $b$ is a learnable bias. Let $y_{ij} \in \{-1, 1\}$ represent the label for the pair $(i, j)$, where $y_{ij}=1$ if $i=j$ (a matched spectro-spatial pair) and $-1$ otherwise. The alignment loss is computed as the sum of independent binary cross-entropy losses across all pairs:
\begin{equation}
    \mathcal{L}_{\text{align}} = - \frac{1}{B^2} \sum_{i=1}^{B} \sum_{j=1}^{B} \log \sigma \left( y_{ij} \cdot s_{ij} \right),
\end{equation}
where $\sigma(\cdot)$ is the sigmoid function. This establishes a grounded spectro-spatial basis for the subsequent generative modeling.

\subsection{State--Space–Conditioned EEG-to-Text Generation}



\subsubsection{Multimodal Large Language Model Architecture}

To bridge the discrete symbolic space of natural language with the continuous manifold of electrophysiology, NeuroNarrator is architected as a unified MLLM, as illustrated in Fig. \ref{fig:model}b. Unlike standard language models that operate exclusively on discrete text tokens, our framework requires a mechanism to ingest dense physiological data directly into the model's input stream. Inspired by the continuous feature injection paradigm recently explored in cardiac modeling\cite{lan2025gem}, we map continuous signals into the MLLM’s high-dimensional embedding space, effectively treating them as ``soft prompts'' that substitute specific placeholders in the instruction sequence.

Our architectural design constructs a composite input sequence that explicitly grounds the generation in both historical temporal dynamics and instantaneous spectro-spatial evidence. Rather than processing segments in isolation, we inject the embeddings of preceding EEG segments to represent the latent trajectory of brain dynamics, followed by the aligned representations of the current segment.

Formally, we utilize the dual-stream encoders and projection heads (defined in Sec. \ref{sec4.2}) to extract aligned embeddings $\mathbf{z}^{\text{eeg}}$ and $\mathbf{z}^{\text{vis}}$ for the EEG segments and topographic maps. Let $\mathcal{M}(\cdot|\theta_{\text{proj}})$ denote the unified embedding function that projects physiological signals into the MLLM’s input space. We construct the multimodal input sequence, $E_{\text{in}}$, by fusing the historical state-space trajectory with the current spectro-spatial observations:
\begin{equation}
E_{\text{in}} = \left[ \underbrace{\mathcal{M}(x_{t-N}\mid \theta_{\text{proj}}^{\text{eeg}}), \dots, \mathcal{M}(x_{t-1}\mid \theta_{\text{proj}}^{\text{eeg}})}_{\text{Historical Context}}, \underbrace{\mathcal{M}(x_{t}\mid \theta_{\text{proj}}^{\text{eeg}}), \mathcal{M}(i_{t}\mid \theta_{\text{proj}}^{\text{vis}})}_{\text{Current Spectro-Spatial Evidence}}, E_{\text{instr}} \right]
\end{equation}
where the sequence $\{x_{t-k}\}_{k=1}^{N}$ represents the temporal context encoded solely via the EEG branch to efficiently model state evolution, while the current target is represented by the concatenation of both its temporal waveform embedding, $\mathcal{M}(x_{t})$, and its spatial topographic embedding, $\mathcal{M}(i_{t})$. $E_{\text{instr}}$ denotes the sequence of token embeddings corresponding to the textual task instruction. This fully constructed sequence $E_{\text{in}}$ is subsequently fed into the Large Language Model (LLM) backbone, enabling the model to jointly process the interleaved physiological embeddings and textual instructions to autoregressively synthesize the final clinical description.

\subsubsection{Instruction-Guided End-to-End Optimization}
 The generation process is explicitly conditioned on two complementary instruction components appended to the unified spectro-spatial sequence $E_{in}$: a state-space prior instruction that contextualizes the current segment with respect to its preceding EEG trajectory, and a task instruction that specifies the requirement for clinically grounded, structured narrative synthesis. This instruction prompts the MLLM to synthesize the latent physiological trajectory into a coherent clinical narrative. A selective optimization strategy is employed during training: the CLIP vision encoder remains frozen to maintain its pretrained semantic alignment, while the EEG encoder, projection layers, and the LLM backbone are jointly updated. The entire framework is optimized end-to-end by minimizing the negative log-likelihood of the target description sequence $y = (y_1, \dots, y_L)$ of length $L$:
\begin{equation}
    \mathcal{L}(\Theta) = -\sum_{l=1}^{L} \log p_\Theta(y_l \mid y_{<l}, E_{in}),
\end{equation}
where $p_\Theta$ denotes the conditional probability of the $l$-th token predicted by the model parameters $\Theta$, which encompass the EEG encoder, projectors, and the language backbone. Under this objective, the extracted spectro-spatial features are grounded within the linguistic domain, adapting to the semantic and structural constraints imposed by the clinical instructions.

\section{Experimental Setup}
\subsection{Model Configuration} In NeuroNarrator, EEG segments are encoded using LaBraM-Base\cite{jiang2024large}, while corresponding scalp topographic maps are processed by a CLIP ViT-Large-Patch14\cite{radford2021learning} vision encoder. The language backbone is instantiated as Qwen3-4B-Instruct\cite{yang2025qwen3}, which serves as the generative core for clinically grounded text synthesis. Temporal context is incorporated by conditioning generation on the embeddings of the preceding $1 \le N \le 3$ EEG segments. To inject multimodal information directly into the prompt as continuous embeddings, the EEG and vision projection heads are configured to output vectors with the same dimensionality as the LLM hidden size (2560).

\subsection{Training Protocol}
All experiments were conducted on a multi-GPU server equipped with four NVIDIA H100 GPUs, using Python 3.11.14 and PyTorch 2.9.1 with CUDA 12.8. In the contrastive spectro–spatial alignment stage, the model was trained for 60 epochs with a per-GPU batch size of 32 and gradient accumulation over four steps, optimized using AdamW\cite{loshchilov2017decoupled} with a learning rate of $5 \times 10^{-5}$. In the unified spectro–spatial interpretation and clinical event generation stage, training was performed for four epochs with a per-GPU batch size of 3 and the same gradient accumulation strategy. All trainable components were jointly optimized using AdamW\cite{loshchilov2017decoupled} with a learning rate of $1 \times 10^{-4}$. To improve computational efficiency and scalability, DeepSpeed\cite{rasley2020deepspeed}, FlashAttention\cite{dao2022flashattention}, and mixed-precision training\cite{micikevicius2017mixed} with bfloat16\cite{wang2019bfloat16} were employed throughout all experiments.

\subsection{Datasets and Evaluation Splits} \label{sec_datasets}
\noindent \textbf{Data Split and Sampling Protocol.} We aligned the subject-level partitioning for TUAB\cite{harati2015improved}, TUEV\cite{harati2015improved}, SEED\cite{zheng2015investigating}, HMC\cite{alvarez2021inter}, Workload\cite{zyma2019electroencephalograms}, and TUSL\cite{von2017electroencephalographic} strictly with the protocols established in NeuroLM\cite{jiang2024neurolm}, while employing random subject-wise splitting for the remaining heterogeneous datasets. To ensure computational tractability and class balance, we restricted each training partition to approximately 20,000 segments, employing a stratified strategy that prioritizes the retention of pathological events over dominant background activity.

\noindent \textbf{Training Data.} All experiments are conducted using the training split of NeuroCorpus-160K. During training, each EEG segment is paired with its corresponding scalp topographic map, rendered at the temporal median index $\tau = \lfloor T/2 \rfloor$ using the MNE-Python toolbox\cite{gramfort2013meg}.

\section{Experimental Results}



\subsection{Validation of Spectro-Spatial Grounding}

\begin{figure}[t]
  \centering
  \includegraphics[width=\textwidth]{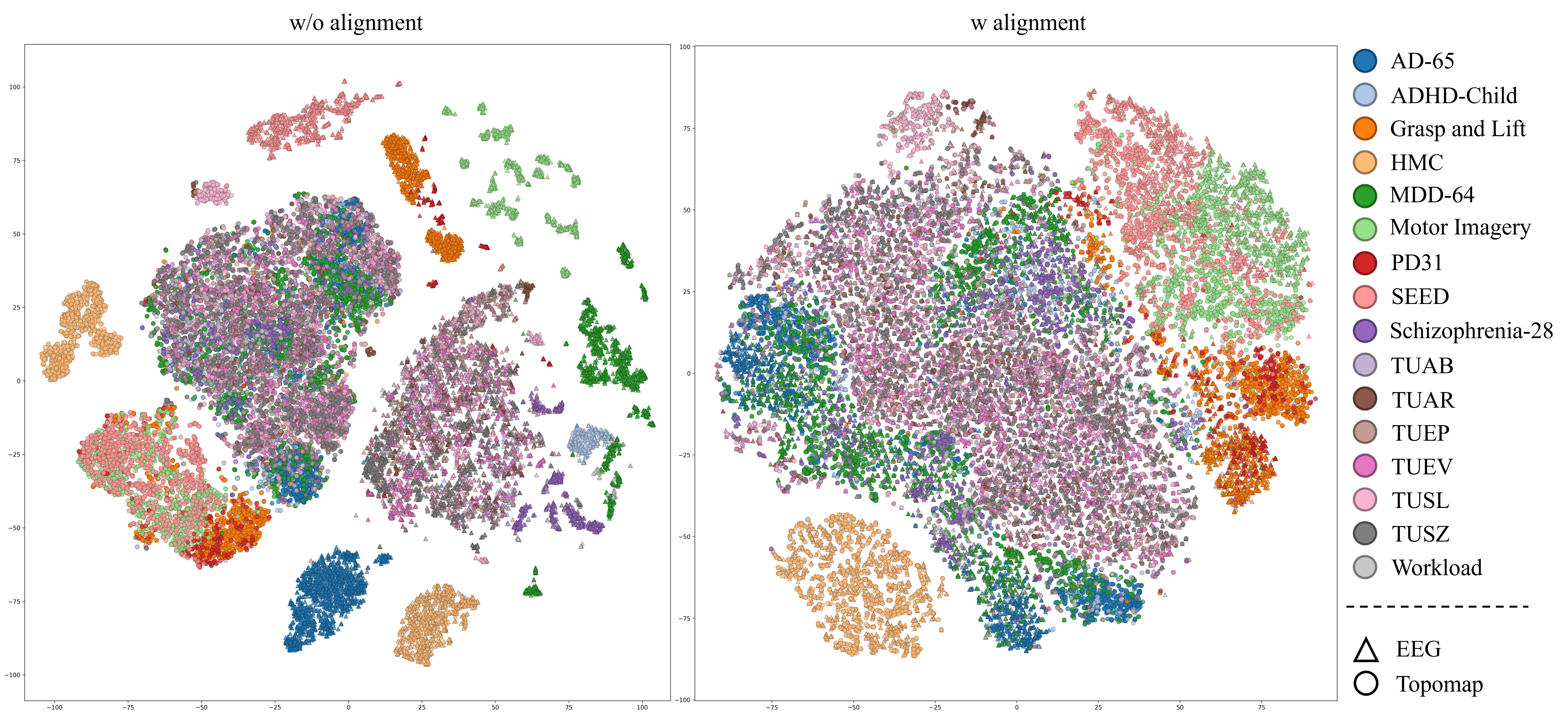}%
\caption{\textbf{Visualization of the learned spectro-spatial manifold via t-SNE\cite{maaten2008visualizing} projection.} The plots depict the latent distribution of EEG segments (triangles) and corresponding topographic maps (circles) sampled from the held-out NeuroCorpus-160K evaluation split. Left: In the absence of contrastive alignment, the representations exhibit a distinct modality gap, with temporal and spatial features forming fragmented, disjoint clusters. Right: Following contrastive optimization, the embedding space demonstrates a rigorous topological correspondence; matched EEG and topographic map pairs are tightly co-located within dataset-specific clusters.}
  \label{fig:alignment}
\end{figure}

\begin{figure}[t]
  \centering
  \includegraphics[width=\textwidth]{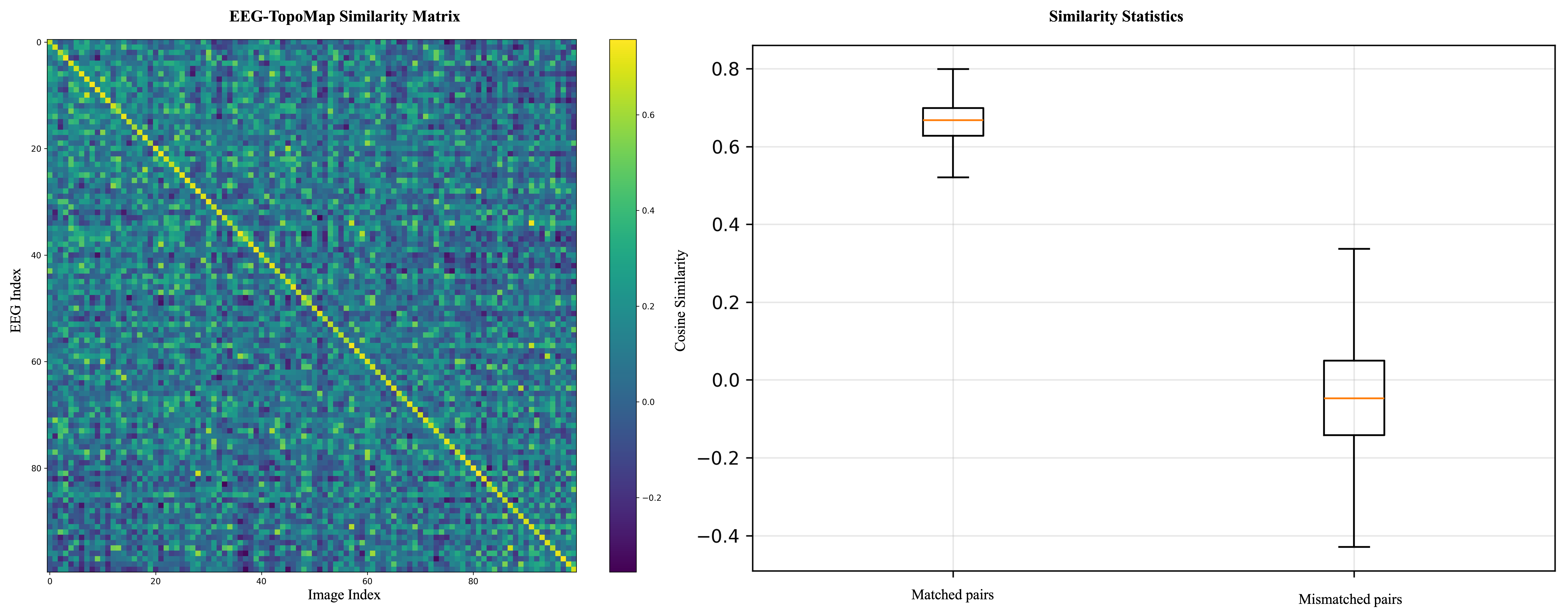}%
\caption{\textbf{Statistical validation of the learned spectro-spatial metric space.} Left: Cosine similarity matrix. The pronounced diagonal dominance confirms a rigorous one-to-one correspondence between temporal EEG embeddings and spatial topographic map embeddings, effectively minimizing off-diagonal ambiguity. Right: Distribution of cosine similarity scores for matched versus mismatched pairs.}

  \label{fig:similarity}
\end{figure}

To evaluate the effectiveness of the proposed contrastive spectro–spatial grounding stage, we conduct cross-modal retrieval experiments on the held-out test split of NeuroCorpus-160K. Following standard evaluation protocols in multimodal alignment, we assess bidirectional retrieval performance between EEG embeddings and topographic map embeddings. We report Recall@K ($R@1$, $R@5$, $R@10$) and Mean Rank ($MeanR$) to measure retrieval precision. As summarized in Table \ref{tab:eeg_image_retrieval}, the contrastive spectro-spatial alignment mechanism achieves robust performance across the heterogeneous acquisition protocols, yielding an average $R@1$ of 84.19\% for EEG-to-topographic map retrieval and 86.09\% for the inverse direction. These quantitative results are corroborated by visualizations of the learned embedding space (Fig. \ref{fig:alignment} and Fig. \ref{fig:similarity}). 

\begin{table*}[t]
\centering
\caption{\textbf{Quantitative assessment of bidirectional spectro-spatial retrieval.}
The table reports cross-modal retrieval performance across 16 heterogeneous datasets, evaluating the alignment accuracy between temporal EEG embeddings and spatial topographic map embeddings. Metrics include Recall@K ($R@1, R@5, R@10$; higher is better) and Mean Rank ($MeanR$; lower is better). 
}
\label{tab:eeg_image_retrieval}
\scriptsize
\setlength{\tabcolsep}{4pt}
\renewcommand{\arraystretch}{1.12}
\resizebox{\textwidth}{!}{
\begin{tabular}{l r r r r r r r r}
\toprule
\multirow{2}{*}{\textbf{Dataset}} &
\multicolumn{4}{c}{\textbf{Topographic Map Retrieval}} &
\multicolumn{4}{c}{\textbf{EEG Signal Retrieval}} \\
\cmidrule(lr){2-5}\cmidrule(lr){6-9}
& R@1 & R@5 & R@10 & MeanR & R@1 & R@5 & R@10 & MeanR \\
\midrule
AD-65\cite{ds004504:1.0.2} & 88.35 & 98.91 & 99.71 & 1.38 & 92.04 & 98.99 & 99.78 & 1.22 \\
ADHD-Child\cite{rzfh-zn36-20} & 87.27 & 96.06 & 98.48 & 2.58 & 88.79 & 96.97 & 98.18 & 1.87 \\
Grasp\_and\_Lift\cite{luciw2014multi} & 74.72 & 93.96 & 97.32 & 2.19 & 77.63 & 95.19 & 97.87 & 1.88 \\
MDD-64\cite{Mumtaz2016} & 85.95 & 97.82 & 99.11 & 1.50 & 87.65 & 98.29 & 99.05 & 1.51 \\
Motor Imagery\cite{schalk2004bci2000} & 71.45 & 91.94 & 95.89 & 2.94 & 75.82 & 93.70 & 96.47 & 2.54 \\
PD31\cite{ds002778:1.0.5} & 77.42 & 93.55 & 95.85 & 2.64 & 77.88 & 92.63 & 94.93 & 2.47 \\
SEED\cite{zheng2015investigating} & 63.05 & 84.57 & 91.14 & 9.01 & 64.48 & 87.33 & 93.05 & 8.50 \\
Schizophrenia-28\cite{repod.0107441_2017} & 92.50 & 98.82 & 99.85 & 1.19 & 94.71 & 99.26 & 100.00 & 1.12 \\
TUAB\cite{harati2015improved} & 95.90 & 99.50 & 99.80 & 1.12 & 96.50 & 99.60 & 99.90 & 1.09 \\
TUAR\cite{buckwalter2021recent} & 87.75 & 94.80 & 96.30 & 9.60 & 87.75 & 94.91 & 96.42 & 9.41 \\
TUEP\cite{veloso2017big} & 85.60 & 95.80 & 97.60 & 2.40 & 88.00 & 96.80 & 98.20 & 1.86 \\
TUEV\cite{harati2015improved} & 91.40 & 98.70 & 99.50 & 1.31 & 92.30 & 98.80 & 99.40 & 1.28 \\
TUSL\cite{von2017electroencephalographic} & 89.44 & 98.91 & 99.64 & 1.53 & 91.44 & 99.09 & 99.64 & 1.58 \\
TUSZ\cite{shah2018temple} & 91.00 & 97.40 & 98.30 & 2.43 & 91.70 & 97.20 & 98.40 & 2.33 \\
Workload\cite{zyma2019electroencephalograms} & 94.44 & 98.89 & 100.00 & 1.16 & 96.67 & 98.89 & 98.89 & 1.13 \\
HMC\cite{alvarez2021inter} & 82.50 & 98.20 & 99.60 & 1.36 & 83.70 & 98.90 & 99.90 & 1.31 \\
\midrule
\textbf{Overall} & 84.19 & 95.99 & 97.93 & 2.85 & 86.09 & 96.66 & 98.22 & 2.66 \\
\bottomrule
\end{tabular}
}
\end{table*}




\subsection{Fidelity of Clinical Narrative Generation} \label{lab:sec6.2}
To assess the model's ability to translate electrophysiological dynamics into clinically accurate narratives, we evaluated NeuroNarrator on the held-out test split of NeuroCorpus-160K. During inference, we employed the same unified task instruction used during the supervised fine-tuning stage to ensure distributional consistency.

We first examine standard text-generation metrics to quantify the lexical and semantic alignment between the generated descriptions and ground-truth reports. As summarized in Table \ref{tab:evaluation_results}, NeuroNarrator achieves an average BERTScore\cite{zhang2019bertscore} of $0.731$, indicating a strong preservation of semantic content despite the inherent variability of natural language expression. The rule-based Fact-F1\cite{yang2018hotpotqa} scores provide a measure of factual consistency, confirming that the model minimizes hallucinations and remains faithful to the underlying signal data. Additionally, the ROUGE-L\cite{lin2003automatic} scores—while naturally lower due to the diverse phrasing styles in clinical reporting—demonstrate sufficient structural overlap to support the validity of the generated syntax.

\begin{figure}[t]
  \centering
  \includegraphics[width=\textwidth]{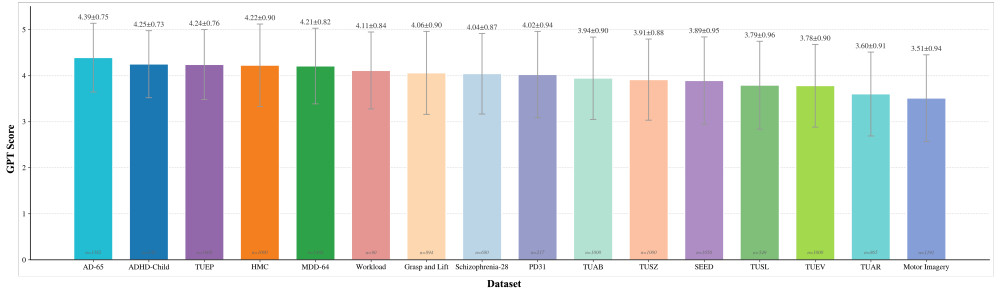}%
\caption{\textbf{Multi-dimensional clinical evaluation of generated narratives.} We assess the physiological fidelity of NeuroNarrator on the held-out NeuroCorpus-160K test set using a structured GPT-4.1 adjudicator. To capture the complexity of expert interpretation, each generated narrative is scored against the ground truth along five complementary axes: (i) clinical event or condition identification, (ii) dominant brain region localization, (iii) dominant frequency band classification, (iv) temporal trend description, and (v) detection of salient non-dominant spectral changes, with each axis contributing one point to a maximum score of 5.
}
\label{fig:gpt_socre}
\end{figure}

While lexical metrics provide necessary quantitative baselines, they often fail to capture the physiological correctness required for clinical utility. To address this, we employed a multi-dimensional evaluation protocol utilizing GPT-4.1\cite{achiam2023gpt} as a structured adjudicator. This framework assesses each generated narrative along five axes critical to EEG interpretation: (i) Clinical Event Identification: identification of the clinical event or condition, (ii) Anatomical Localization: anatomical localization of the dominant signal, (iii) Dominant Frequency Classification: classification of the dominant frequency band, (iv) Dominant Trend Characterization: description of the primary evolving dynamics, and (v) Non-Dominant Spectral Detection: detection of salient changes in non-dominant bands, with each axis contributing one point to a maximum score of 5 (Fig. \ref{fig:gpt_socre}). 

\begin{figure}[t]
  \centering
  \includegraphics[width=0.6\textwidth]{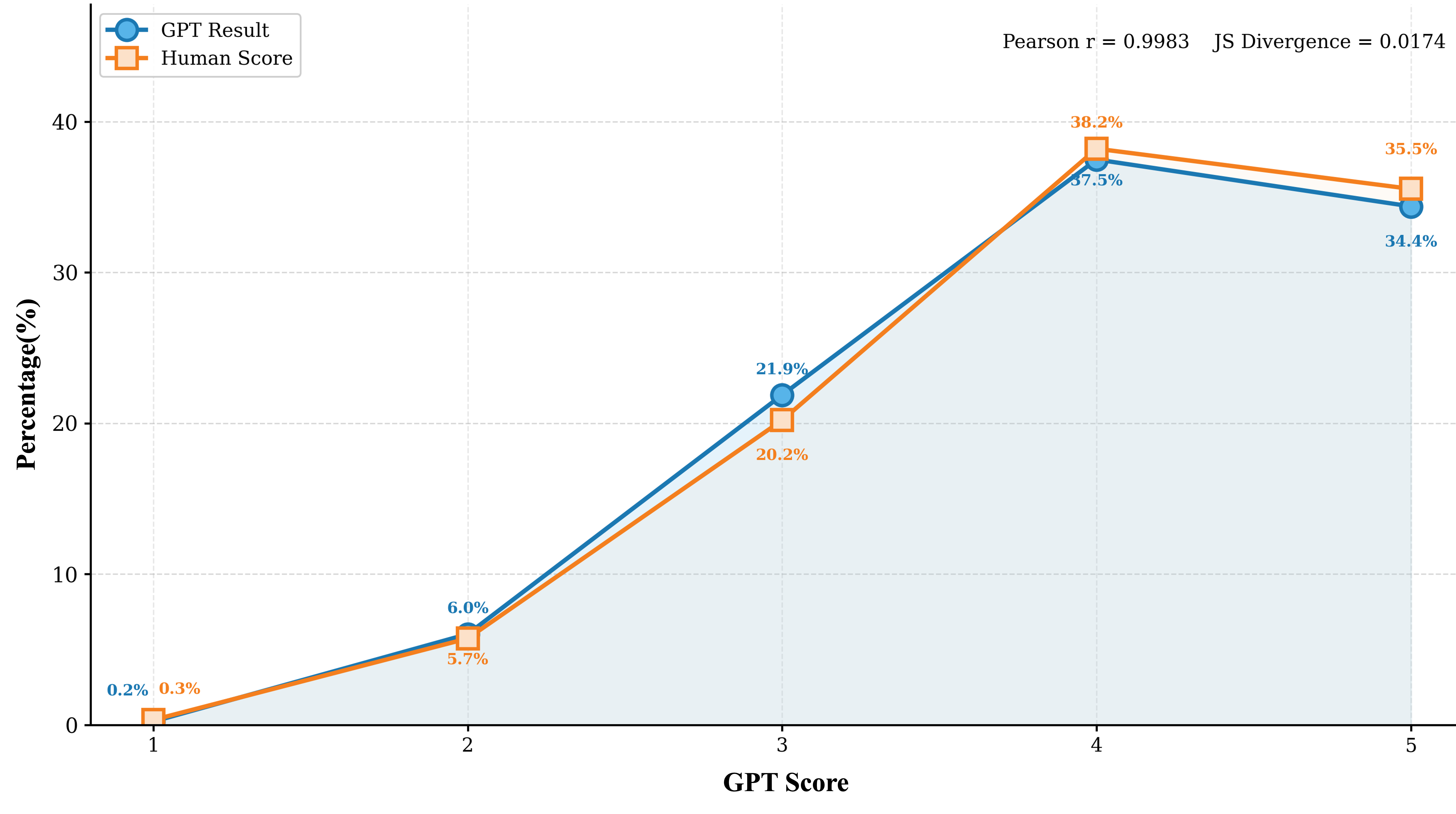}%
\caption{\textbf{Alignment of human and automated scoring distributions.} Comparison of clinical fidelity scores assigned by human experts versus the GPT-4.1 adjudicator on a stratified subset of 1,500 narratives.}
\label{fig:gpt_credibility}
\end{figure}

To validate the reliability of GPT-4.1\cite{achiam2023gpt} as a proxy for clinical judgment, we conducted a human study on a random subset of 1,500 generated captions. Independent expert scoring yielded a distribution that closely mirrors that of the automated evaluator (Fig. \ref{fig:gpt_credibility}). 


\begin{table*}[htbp]
\centering
\caption{\textbf{Quantitative assessment of linguistic and semantic fidelity on NeuroCorpus-160K.} Performance of NeuroNarrator across all held-out benchmark datasets is evaluated using complementary automated metrics: BERTScore, Rule-based Fact-F1 and ROUGE-L.}
\label{tab:evaluation_results}
\resizebox{\textwidth}{!}{
\begin{tabular}{l|ccc}
\toprule
\textbf{Dataset} & \textbf{BERTScore} & \textbf{Fact-F1} & \textbf{ROUGE-L} \\
\midrule
AD-65\cite{ds004504:1.0.2} & 0.750 $\pm$ 0.045 & 0.665 $\pm$ 0.237 & 0.419 $\pm$ 0.079 \\
ADHD-Child\cite{rzfh-zn36-20} & 0.743 $\pm$ 0.049 & 0.706 $\pm$ 0.218 & 0.411 $\pm$ 0.084 \\
Grasp and Lift\cite{luciw2014multi} & 0.728 $\pm$ 0.044 & 0.737 $\pm$ 0.251 & 0.365 $\pm$ 0.078 \\
MDD-64\cite{Mumtaz2016} & 0.732 $\pm$ 0.044 & 0.747 $\pm$ 0.209 & 0.386 $\pm$ 0.079 \\
Motor Imagery\cite{schalk2004bci2000} & 0.718 $\pm$ 0.044 & 0.692 $\pm$ 0.247 & 0.353 $\pm$ 0.076 \\
PD-31\cite{ds002778:1.0.5} & 0.741 $\pm$ 0.053 & 0.683 $\pm$ 0.235 & 0.397 $\pm$ 0.091 \\
SEED\cite{zheng2015investigating} & 0.734 $\pm$ 0.041 & 0.741 $\pm$ 0.252 & 0.385 $\pm$ 0.076 \\
Schizophrenia-28\cite{repod.0107441_2017} & 0.739 $\pm$ 0.048 & 0.629 $\pm$ 0.217 & 0.397 $\pm$ 0.089 \\
TUAB\cite{harati2015improved} & 0.734 $\pm$ 0.047 & 0.736 $\pm$ 0.241 & 0.377 $\pm$ 0.083 \\
TUAR\cite{buckwalter2021recent} & 0.725 $\pm$ 0.049 & 0.672 $\pm$ 0.267 & 0.365 $\pm$ 0.084 \\
TUEP\cite{veloso2017big} & 0.712 $\pm$ 0.047 & 0.665 $\pm$ 0.220 & 0.343 $\pm$ 0.079 \\
TUEV\cite{harati2015improved} & 0.719 $\pm$ 0.050 & 0.694 $\pm$ 0.245 & 0.359 $\pm$ 0.085 \\
TUSL\cite{von2017electroencephalographic} & 0.718 $\pm$ 0.047 & 0.653 $\pm$ 0.255 & 0.361 $\pm$ 0.079 \\
TUSZ\cite{shah2018temple} & 0.722 $\pm$ 0.050 & 0.691 $\pm$ 0.233 & 0.363 $\pm$ 0.086 \\
Workload\cite{zyma2019electroencephalograms} & 0.741 $\pm$ 0.044 & 0.741 $\pm$ 0.250 & 0.378 $\pm$ 0.083 \\
HMC\cite{alvarez2021inter} & 0.755 $\pm$ 0.050 & 0.757 $\pm$ 0.235 & 0.423 $\pm$ 0.094 \\
\midrule
\textbf{Average} & \textbf{0.731 $\pm$ 0.048} & \textbf{0.703 $\pm$ 0.241} & \textbf{0.379 $\pm$ 0.086} \\
\bottomrule
\end{tabular}
}
\end{table*}



\subsection{Comparison with Specialized Baselines} 

\begin{table}[t] \centering \caption{\textbf{Benchmarking Open-Vocabulary Interpretation Against Specialized Closed-Set Baselines.} We report Balanced Accuracy across six representative subsets of NeuroCorpus-160K. The evaluation contrasts NeuroNarrator (an open-vocabulary generative model) against two categories of discriminative baselines: conventional non-LLM deep learning models and NeuroLM, a closed-set LLM baseline.} \label{tab:sota} \begin{tabular}{l c c c c c c} \toprule \textbf{Methods} & \textbf{TUAB} & \textbf{TUEV} & \textbf{TUSL} & \textbf{SEED} & \textbf{HMC} & \textbf{Workload} \\ \midrule \multicolumn{7}{c}{Non-LLM Models} \\ \midrule EEGNet\cite{lawhern2018eegnet} & 79.3{\scriptsize$\pm$0.3} & 35.7{\scriptsize$\pm$1.0} & 36.1{\scriptsize$\pm$3.8} & 47.1{\scriptsize$\pm$0.8} & 62.2{\scriptsize$\pm$0.8} & 61.1{\scriptsize$\pm$5.5} \\ EEGConformer\cite{song2022eeg} & 80.1{\scriptsize$\pm$0.3} & 32.1{\scriptsize$\pm$1.0} & 44.9{\scriptsize$\pm$2.1} & 44.9{\scriptsize$\pm$0.8} & 49.5{\scriptsize$\pm$0.8} & 75.5{\scriptsize$\pm$4.0} \\ SPaRCNet\cite{jing2023development} & 79.8{\scriptsize$\pm$1.1} & 31.4{\scriptsize$\pm$3.6} & 40.0{\scriptsize$\pm$3.2} & 48.6{\scriptsize$\pm$2.4} & 66.0{\scriptsize$\pm$1.7} & 73.2{\scriptsize$\pm$4.9} \\ ContraWR\cite{yang2023self} & 82.3{\scriptsize$\pm$0.3} & 42.2{\scriptsize$\pm$1.4} & 39.9{\scriptsize$\pm$1.4} & 56.4{\scriptsize$\pm$2.8} & 69.8{\scriptsize$\pm$0.8} & 63.3{\scriptsize$\pm$3.1} \\ CNN-Transformer\cite{peh2022transformer} & 82.2{\scriptsize$\pm$0.6} & 36.0{\scriptsize$\pm$4.7} & 35.5{\scriptsize$\pm$3.1} & 56.2{\scriptsize$\pm$1.2} & 69.9{\scriptsize$\pm$0.5} & 58.3{\scriptsize$\pm$7.7} \\ FFCL\cite{li2022motor} & 82.0{\scriptsize$\pm$0.3} & 34.5{\scriptsize$\pm$3.7} & 45.2{\scriptsize$\pm$3.2} & 48.7{\scriptsize$\pm$1.5} & 65.0{\scriptsize$\pm$0.7} & 59.6{\scriptsize$\pm$8.3} \\ ST-Transformer\cite{song2021transformer} & 80.5{\scriptsize$\pm$0.8} & 30.2{\scriptsize$\pm$1.9} & 36.5{\scriptsize$\pm$4.6} & 42.0{\scriptsize$\pm$2.0} & 58.9{\scriptsize$\pm$0.8} & 66.2{\scriptsize$\pm$3.7} \\ BIOT\cite{yang2023biot} & 81.3{\scriptsize$\pm$0.3} & 41.8{\scriptsize$\pm$1.6} & 44.0{\scriptsize$\pm$1.4} & 62.4{\scriptsize$\pm$0.8} & 64.7{\scriptsize$\pm$0.8} & 68.0{\scriptsize$\pm$1.0} \\ LaBraM-Base\cite{jiang2024large} & 83.5{\scriptsize$\pm$0.5} & 37.7{\scriptsize$\pm$1.9} & 40.0{\scriptsize$\pm$5.0} & 57.6{\scriptsize$\pm$1.7} & 63.6{\scriptsize$\pm$1.3} & 68.2{\scriptsize$\pm$3.2} \\ \midrule \multicolumn{7}{c}{Closed-Set LLM Baselines} \\ \midrule NeuroLM-B\cite{jiang2024neurolm} & 70.3{\scriptsize$\pm$0.3} & 23.3{\scriptsize$\pm$1.8} & 39.6{\scriptsize$\pm$1.0} & 44.3{\scriptsize$\pm$0.8} & 64.8{\scriptsize$\pm$0.8} & 71.1{\scriptsize$\pm$1.5} \\ \midrule \multicolumn{7}{c}{Open-Vocabulary Generative Model} \\ \midrule NeuroNarrator & 70.8{\scriptsize$\pm$0.4} & 19.2{\scriptsize$\pm$1.5} & 31.5{\scriptsize$\pm$2.0} & 61.7{\scriptsize$\pm$0.9} & 75.0{\scriptsize$\pm$0.8} & 58.4{\scriptsize$\pm$1.1} \\ \bottomrule \end{tabular} \end{table}

To benchmark NeuroNarrator against established discriminative frameworks, we evaluated performance on six representative data subsets derived from NeuroCorpus-160K: TUAB\cite{harati2015improved}, TUEV\cite{harati2015improved}, SEED\cite{zheng2015investigating}, HMC\cite{alvarez2021inter}, TUSL\cite{von2017electroencephalographic}, and Workload\cite{zyma2019electroencephalograms}. To ensure a strictly controlled comparison, all baseline models were trained and evaluated exclusively on the specific subsets integrated into the NeuroCorpus-160K training and held-out test splits, rather than on the complete original public repositories. Experimental protocols differ fundamentally between our generalist framework and the baselines: while all baseline models were individually fine-tuned and optimized specifically for each downstream dataset to maximize task-specific performance, NeuroNarrator was evaluated as a single, unified model across all benchmarks without any dataset-specific fine-tuning or further parameter adaptation.

Since NeuroNarrator is designed as an open-vocabulary generative model rather than a closed-set classifier, it does not output probability distributions over fixed class labels. Consequently, we employed deterministic rule-based parsing to extract task-specific predictions from the generated text, specifically targeting the ``Event \& Clinical Labels'' component. Notably, direct comparisons between NeuroNarrator and these baselines are not entirely fair, as the baseline methods are explicitly trained and optimized for closed-set classification on individual datasets. We compared NeuroNarrator against two categories of baselines as detailed in Table \ref{tab:sota}: (i) conventional closed-set deep learning models that operate without language backbones; and (ii) NeuroLM\cite{jiang2024neurolm}, which leverages a large language model but remains architecturally constrained to fixed-label classification rather than open-ended generation.

\subsection{Cross-Domain and Zero-Shot Generalization}
\begin{figure}[t]
  \centering
  \includegraphics[width=1.0\textwidth]{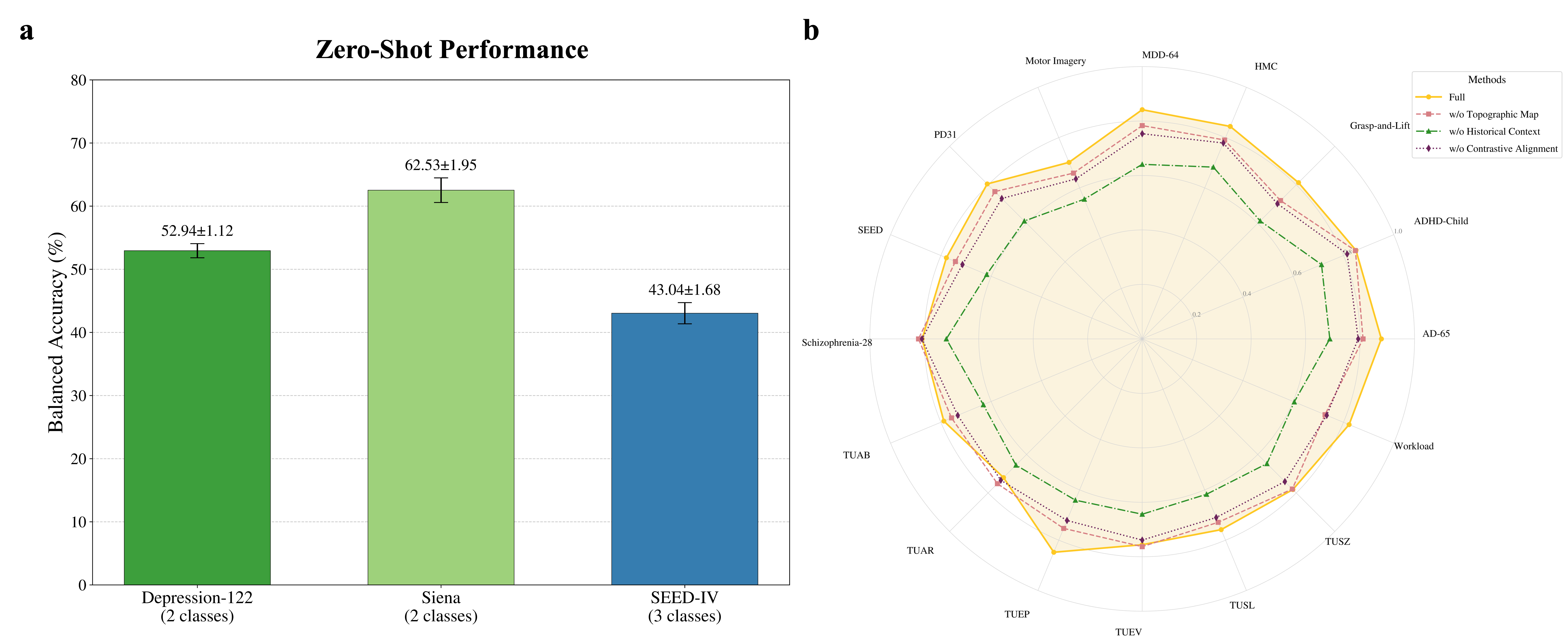}%
\caption{\textbf{Cross-domain zero-shot generalization and architectural ablation analysis.} (a) Zero-shot classification performance on three external datasets (Depression-122, Siena, and SEED-IV) strictly excluded from NeuroCorpus-160K. (b) Ablation study on NeuroCorpus-160K quantifying the contribution of specific modules to clinical narrative fidelity.}
\label{fig:zero_shot_ablation}
\end{figure}

To assess generalization capabilities beyond the training distribution, we subjected NeuroNarrator to a zero-shot inference protocol on three external datasets strictly excluded from the NeuroCorpus-160K curation: Depression-122 (Major Depressive Disorder Detection)\cite{ds003478:1.1.0}, Siena Scalp EEG (Epilepsy Detection)\cite{detti2020eeg}, and SEED-IV (Emotion Recognition)\cite{8283814}. All external recordings were processed via the fixed preprocessing pipeline detailed in Sections \ref{lab:sec3.2} and \ref{lab:sec3.3}, with the model parameters kept strictly frozen to preclude any parameter adaptation or supervised fine-tuning. Following label space harmonization for SEED-IV—where `sad' and `fear' were aggregated into a unified negative class to align with the three-class label space (Positive, Neutral, Negative) of the training source—NeuroNarrator achieved balanced accuracies of 52.94\% (Depression-122), 62.53\% (Siena), and 43.04\% (SEED-IV) (Fig. \ref{fig:zero_shot_ablation}a), demonstrating the capability to generalize to unseen distributions despite the rigorous domain shift.


\subsection{Ablation Study}
To validate the architectural efficacy of NeuroNarrator, we conducted ablation studies on NeuroCorpus-160K. Performance was assessed using the structured GPT-4.1\cite{achiam2023gpt} adjudication protocol established in Section \ref{lab:sec6.2}, which quantifies clinical narrative fidelity on a maximum scale of 5. We systematically evaluated the model by isolating specific components, comparing the full framework against three ablated configurations consistent with Figure \ref{fig:zero_shot_ablation}b: (i) w/o Contrastive Alignment, where the spectro-spatial alignment objective is omitted; (ii) w/o Topographic Map, which excludes the vision-encoded spatial stream; and (iii) w/o Historical Context, where the historical state-space conditioning is removed. As visualized in Fig. \ref{fig:zero_shot_ablation}b, the Full Model systematically outperforms all ablated variants, confirming the essential contribution of each module to the overall interpretative fidelity.


\section{Discussion}
This work introduces NeuroNarrator, a generalist framework for translating electrophysiological recordings into segment-level, clinically grounded natural-language descriptions, addressing the fundamental challenge of bridging continuous neural dynamics and discrete clinical language in EEG interpretation. A key enabling factor is our construction of NeuroCorpus-160K, the first large-scale EEG–text dataset that provides temporally localized signal–narrative pairings, enabling supervision at the segment level and allowing models to learn signal-grounded and temporally coherent clinical narratives rather than coarse recording-level summaries. By explicitly aligning the three intrinsic dimensions of EEG—temporal dynamics, spectral structure, and spatial topology—with the semantic latent space of a MLLM, our results demonstrate that deep generative architectures can internalize and operate upon the compositional logic underlying expert EEG interpretation. In contrast to conventional paradigms that compress rich neurophysiological phenomena into fixed categorical labels, NeuroNarrator advances a shift toward signal-grounded, white-box interpretation, showing that multimodal language models can synthesize heterogeneous electrophysiological evidence into coherent narratives that reflect the evolving state of the brain rather than static diagnostic endpoints.

These conceptual advances are further supported by the experimental results, which provide several insights into the strengths of the proposed framework. Across diverse datasets spanning pathological electrophysiological events, cognitive states, and artifact conditions, NeuroNarrator demonstrates consistently strong performance without dataset-specific fine-tuning, highlighting the robustness of the learned representations under substantial distributional heterogeneity. Notably, the model maintains stable narrative fidelity even in settings where conventional classifiers are explicitly optimized for label accuracy, underscoring the advantage of a unified, open-vocabulary formulation over task-specific optimization. The ablation analyses further clarify the contributions of individual architectural components, showing that removing historical temporal context leads to the most consistent performance degradation across datasets, underscoring the central role of temporally conditioned generation in capturing non-stationary EEG dynamics. From a neurophysiological perspective, EEG signals are inherently non-stationary, and clinically meaningful patterns are often expressed through temporally evolving spectral structure rather than isolated snapshots, with cognitive and pathological states encoded in the progression and modulation of oscillatory activity over time~\cite{makeig2002dynamic,breakspear2017dynamic}. Incorporating historical context therefore facilitates the modeling of latent neural dynamics and supports the extraction of semantically meaningful patterns aligned with clinical interpretation. This finding is consistent with the state-space-inspired design of NeuroNarrator, which jointly models temporally evolving spectral and spatial structure to support coherent narrative interpretation.

Despite achieving consistently high scores under the structured GPT-based clinical adjudication, NeuroNarrator exhibits comparatively lower precision on ``Event \& Clinical Label'' extraction on certain datasets. This discrepancy, however, should not be interpreted as a limitation of the model’s interpretative capacity. Unlike closed-set baselines that are individually fine-tuned and explicitly optimized for discrete label prediction on each benchmark, NeuroNarrator is evaluated as a single, frozen generalist model without any dataset-specific adaptation. In this setting, event labels constitute only one facet of a substantially richer descriptive space, which additionally encodes spatial localization, spectral composition, and temporally evolving dynamics. Evaluations restricted to downstream classification accuracy therefore conflate representational fidelity with label conformity under fundamentally different optimization regimes, resulting in an inherently asymmetric comparison that underestimates the expressive scope of open-vocabulary generative EEG interpretation.

This misalignment further highlights the absence of principled evaluation metrics tailored to open-ended physiological description. Current practices, whether label-centric benchmarks or LLM-based textual scoring, remain imperfect proxies for assessing how faithfully a model captures clinically meaningful signal attributes. Developing physiology-aware, modality-specific evaluation frameworks thus represents a critical direction for future work. Similarly, zero-shot generalization in the present setting reflects an early-stage scaling regime. Given the frozen-parameter protocol and the magnitude of distributional shift, the consistent performance indicates that NeuroNarrator already internalizes transferable electrophysiological abstractions, which are expected to strengthen substantially with increased pretraining diversity and model capacity.

Looking ahead, this work suggests a broader research direction in which EEG interpretation is formulated as a problem of signal-grounded language generation rather than discrete label prediction. As larger and more diverse electrophysiological corpora become available, such formulations may naturally extend to richer temporal reasoning, integration of additional clinical context, and more flexible forms of cross-condition analysis. In this sense, segment-level EEG-to-text modeling provides a scalable and extensible framework for developing interpretable computational tools that support both offline analysis and real-time, continuously updated interpretation in clinical and research settings.

\section*{Data Availability}
All data used in this study are derived from publicly available sources. The processed dataset (NeuroCorpus-160K) and code supporting the findings of this study will be made available upon acceptance of the manuscript and are available from the corresponding author upon reasonable request.

\section*{Declaration of Interests}
The authors declare no competing interests.

\section*{Acknowledgments}
Research reported in this publication was supported by the National Institute of Neurological Disorders and Stroke (NINDS) of the National Institutes of Health (NIH), United States under Award Number R21NS135482 (PI: Liu). The content is solely the responsibility of the authors and does not necessarily represent the official views of the National Institutes of Health. 
This work utilized computational resources provided by the National Artificial Intelligence Research Resource (NAIRR) Pilot Program under grant number NAIRR240428 through the Anvil supercomputer at Purdue University. The authors gratefully acknowledge NAIRR and Anvil for the computing support that enabled the large-scale experiments presented in this study.

\begin{appendices}






\end{appendices}


\bibliography{sn-bibliography}

\end{document}